\documentclass[preprintnumbers,prd,aps,nofootinbib,twocolumn,showpacs,amsmath,superscriptaddress]{revtex4-2}
\usepackage{graphicx}
\usepackage{epstopdf}
\usepackage{bm}
\usepackage{epsfig}
\usepackage{float}
\usepackage{dblfloatfix}
\usepackage{graphics}
\usepackage{xspace}
\usepackage{caption}
\usepackage{subcaption}
\usepackage{amsmath}
\usepackage{xcolor}
\usepackage{color}

\usepackage{amssymb}
\usepackage{comment}

\def\OMIT#1{}

\DeclareMathAlphabet{\mymathbb}{U}{BOONDOX-ds}{m}{n}

\newcommand{\nn}{\nonumber}

\newcommand{\bea}{\begin{eqnarray}}
\newcommand{\eea}{\end{eqnarray}}

\newcommand{\gsim}{{\cal M}threl{\rlap{\lower4pt\hbox{\hskip1pt$\sim$}}\raise1pt\hbox{$>$}}}

\newcommand{\be}{\begin{equation}}
\newcommand{\ee}{\end{equation}}

\allowdisplaybreaks

\begin{document}

\preprint{JLAB-THY-26-4823
}

\title{\bf Impact of QED Radiation on SMEFT Constraints in  Deep Inelastic Scattering}

\author{Sonny Mantry}
\affiliation{Department of Physics and Astronomy, 
                   University of North Georgia,
                   Dahlonega, GA 30597, USA}

\author{Jian-Wei Qiu}
\affiliation{Department of Physics, William \& Mary, Williamsburg, VA 23187, USA}
\affiliation{Theory Center, Jefferson Lab, Newport News, VA, 23606, USA}

\author{Jia-Yue Zhang}
\affiliation{Theory Center, Jefferson Lab, Newport News, VA, 23606, USA}

\begin{abstract}
  \vspace*{0.3cm}

  Deep-inelastic scattering (DIS) provides a powerful probe of physics beyond the Standard Model through precision measurements interpreted within the Standard Model Effective Field Theory (SMEFT). We study the impact of collision-induced
  QED radiation on SMEFT constraints using the joint QCD+QED factorization framework based on lepton distribution and fragmentation functions. QED radiation can substantially modify DIS cross sections and, in some kinematic regions, significantly alter the effective momentum transfer relevant for factorization. We find that while cross sections receive order-one corrections, longitudinal electron spin asymmetries are affected only at the few-percent level, making them significantly more robust observables for SMEFT studies. Benchmark projections for SoLID and the Electron-Ion Collider are provided to demonstrate the impact of QED radiation for future precision DIS analyses and the extraction of SMEFT constraints.

\end{abstract}

\maketitle

\textit{Introduction}:  Precision deep inelastic scattering (DIS) measurements provide a unique window on physics beyond the Standard Model through the Standard Model Effective (SMEFT)~\cite{Brivio:2017vri,Buchmuller:1985jz,Arzt:1994gp,Grzadkowski:2010es}, complementary to hadron colliders. The proposed SoLID~\cite{JeffersonLabSoLID:2022iod} fixed target experiment at Jefferson Laboratory (JLAB) and the Electron-Ion Collider (EIC)~\cite{AbdulKhalek:2021gbh} at the Brookhaven National Lab (BNL) will provide high luminosity facilities for precision measurements of DIS observables. Projections of SMEFT constraints from SoLID and the EIC were shown~\cite{Alte:2018xgc,RBoughezal2020,RBoughezal2021,Boughezal:2022pmb}  to be competitive and complementary to those derived from the Drell-Yan process at the Large Hadron Collider (LHC), often probing energy scales beyond the direct energy reach of the LHC. The capability of SoLID to polarize the electron beam and the EIC to polarize both the electron and proton beams, facilitates measurements of a wide variety of DIS cross sections and polarization asymmetries that can be used to disentangle the effects of different SMEFT operators. Furthermore, SMEFT analyses at the LHC can be contaminated by non-negligible contributions from dimension eight~\cite{Murphy:2020rsh,Li:2020gnx} operators due to the large energies involved. By contrast, at the EIC and SoLID,  dimension eight SMEFT operator contributions are highly suppressed~\cite{RBoughezal2021}, simplifying the analysis and providing complementary information.  Extensive work has been done to improve SMEFT analyses by including higher order perturbative corrections~\cite{Passarino:2016pzb} and resummation ~\cite{Jenkins:2013zja,Jenkins:2013wua,Alonso:2013hga}, larger data sets for a comprehensive global analysis~\cite{Han:2004az,Pomarol:2013zra,Chen:2013kfa,Ellis:2014dva,Wells:2014pga,Falkowski:2014tna,deBlas:2016ojx,Cirigliano:2016nyn,Biekotter:2018ohn,Grojean:2018dqj,Hartland:2019bjb,Brivio:2019ius}, and more careful analyses of theoretical and experimental uncertainties~\cite{Keilmann:2019cbp,Boughezal:2022pmb}. All of these previous SMEFT studies were done without including the effects of collision-induced QED radiation from the charged leptons and quarks in the process. 

In this letter, for the first time, we address the impact of QED radiation on SMEFT analyses of DIS, and provide benchmark projections for SoLID and the EIC.  We focus on SMEFT analyses in neutral current (NC) lepton-proton deep inelastic scattering (DIS)
\bea
e^-(\ell) + p(P) \to e^-(\ell') + X,
\eea
where $\ell, \ell',$ and $P$ denote the momenta of the initial electron, final scattered electron, and the initial proton, respectively. The initial and final state leptons can radiate logarithmically enhanced energetic and collinear photons that directly affect the momentum transfer delivered to the  proton. It results in a mismatch between the  momentum transfer reconstructed from the incident and final scattered electron, $Q^2=-q^2=-(\ell-\ell')^2$, and the true squared momentum transfer, $\hat{Q}^2=-\hat{q}^2$, delivered to the proton. This mismatch can  threaten the validity of QCD factorization in DIS. In certain kinematic regions, even if the measured $Q^2 \gg \Lambda_{\rm QCD}^2$, the true momentum transfer  can be $\hat{Q}^2\sim \Lambda^2_{\rm QCD}$, taking us out of the DIS regime where higher twist effects and quasielastic  tails can contribute.  
Thus,  QED radiation can significantly affect theoretical predictions and the interpretation of DIS data.

 Traditionally, the effects of collision-induced QED radiation are treated~\cite{RevModPhys.41.205,Bardin:1989vz,Badelek:1994uq,Kripfganz:1990vm,Spiesberger:1994dm,Blumlein:2002fy}  as radiative correction  (RC) factors  added to the born cross section, defined as the cross section without collision-induced QED radiation. In this approach, the extraction of the true Born cross section becomes an inverse problem, requiring measuring all the QED radiation and subtracting the appropriate RC factors. This can be improved by resummation~\cite{Kripfganz:1990vm,PhysRevD.52.4936,Blumlein:2002fy,Afanasev:2001zg} of the logarithmically enhanced RC factors. However, without being able to experimentally account for all the QED radiation, certain RC factors require  measuring the invariant mass of the hadronic final state and Monte Carlo simulations~\cite{Charchula:1994kf,PierreThesis,Kwiatkowski:1990es,Arbuzov:1995id}. Furthermore, infrared divergences are treated by introducing additional cutoff parameters. In this work, we use the recently proposed~\cite{Liu:2020rvc,Liu:2021jfp,Qiu:2024arw,Cammarota:2025jyr} rigorous and systematically improvable  joint QCD+QED factorization framework to treat~\cite{Whitehill:2026swa} the effects of collision-induced QED radiation.
 
 \begin{figure}
 \centering
    \includegraphics[width=\linewidth]{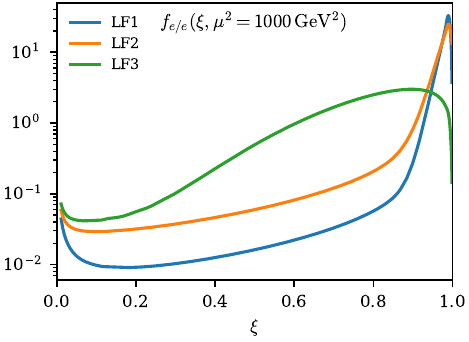}
    \caption{LDF parameterizations used in this analysis.  We consider three different forms labeled as LF1 (blue), LF2 (orange), and LF3 (green). LF1 is chosen as the default parameterization.  }
    \label{fig:ldf}
\end{figure}

\textit{Joint Factorization Approach}:  In this  approach, QED radiation is not treated as an effect that must be subtracted but rather as a contribution to the DIS process that must be included in the same manner as QCD radiation. It treats QED radiation off the initial and final state leptons and the charged quarks on an equal footing with QCD radiation. The joint QCD and QED factorization formula for DIS is schematically given by
\bea
\label{eq:fac}
d\sigma &=& \frac{1}{2S} \sum_{i,j,a} D_{e/j} (\zeta,\mu) \otimes_\zeta \>f_{i/e}(\xi,\mu) \otimes_\xi \>f_{a/h}(x,\mu) \nn \\
&&\otimes_x \>\hat{H}_{ia\to jX}(\hat{s},\hat{t},\hat{u},\mu) ,
\eea
where $d\sigma$ denotes the cross section that may be differential in kinematic variables such as $Q^2$ and the Bjorken variable $x_B$. The indices $i,j,a$ run over all lepton and parton flavors. The symbol $\otimes_t$ for $t=\zeta,\xi$ or $x$ denote convolutions in the momentum fraction of type ``$t$".  The functions $D_{e/j} (\zeta,\mu)$, $f_{i/e}(\xi,\mu)$ and $f_{a/h}(x,\mu)$ denote the lepton fragmentation functions (LFFs), the lepton distribution functions (LDFs), and the parton distribution functions (PDFs), respectively. These distribution functions satisfy DGLAP evolution equations that resum logarithmically enhanced collinear QED and QCD radiation. 
The hard function is given by the partonic cross section, $d\hat{\sigma}(ia\to jX)$, as $\hat{H}_{ia\to jX}(\hat{s},\hat{t},\hat{u},\mu)=2\hat{s} \>d\hat{\sigma}$. It is written as the sum of SM and SMEFT contributions:
\bea
\label{eq:hatH}
\hat{H}_{ia\to jX}(\hat{s},\hat{t},\hat{u},\mu) &=& 2\hat{s} \>\Big \{d\hat{\sigma}^{\gamma \gamma}+ d\hat{\sigma}^{\gamma Z}+ d\hat{\sigma}^{Z Z}\nn \\
&+& d\hat{\sigma}^{\gamma {\rm SMEFT}} +  d\hat{\sigma}^{Z\>{\rm SMEFT}}\Big \},
\eea
corresponding to contributions from single photon exchange, the interference of single photon and Z-boson exchange, Z-boson exchange, the interference of single photon exchange with SMEFT operators, and the interference of single Z-boson exchange with SMEFT operators, respectively. We do not include  contributions arising purely from SMEFT operators since their
contribution scales as $1/\Lambda^4$ in the cross section, the same order as
the interference between dimension-eight SMEFT operators and the SM contributions. 
\begin{figure}
 \centering
    \includegraphics[width=\linewidth]{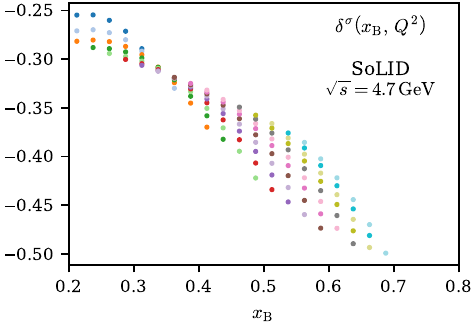}\\
     \includegraphics[width=\linewidth]{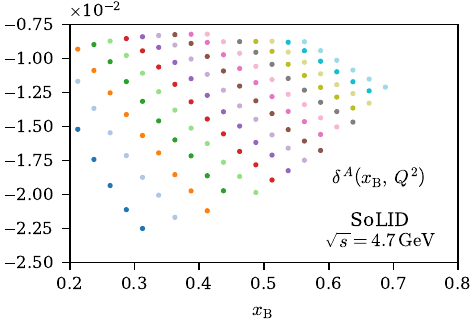}\\
    \caption{Impact of QED radiation on the unpolarized cross section (top) and the longitudinal electron spin asymmetry (bottom) at SoLID, characterized by the deviation $\delta^\sigma(x_B,Q^2)$ and $\delta^{A}(x_B,Q^2)$ in Eq.~(\ref{eq:devsigma}). Each color represents a fixed $Q^2$ bin of width $\Delta Q^2=0.5$ GeV$^2$, covering the range $Q^2= [2,10]$ GeV$^2$.
    }
    \label{fig:deltasigASoLID}
\end{figure}

 \begin{figure}
 \centering
    \includegraphics[width=\linewidth]{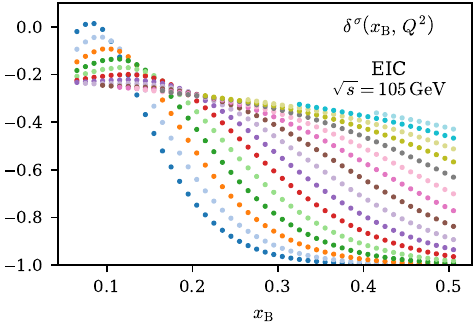}\\
     \includegraphics[width=\linewidth]{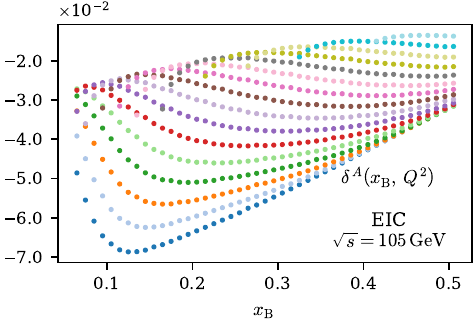}
    \caption{Impact of QED radiation on the unpolarized cross section (top) and the longitudinal electron spin asymmetry (bottom) at the EIC characterized by the deviation $\delta^\sigma(x_B,Q^2)$ and $\delta^{A}(x_B,Q^2)$ in Eq.~(\ref{eq:devsigma}), respectively. Each color represents a fixed $Q^2$ bin of size $\Delta {\rm log}_{10}Q^2=0.15$, in the range  $Q^2= (113,3572)$ GeV$^2$.} 
    \label{fig:deltasigAEIC}
\end{figure}

\textit{Impact of QED Radiation}: In our analysis, we consider the unpolarized cross section
\bea
d\sigma_0 &=& d\sigma_R+d\sigma_L,
\eea
where $d\sigma_R$ and $d\sigma_L$ denote the cross sections with right-handed and left-handed electrons, respectively. The polarized cross section is defined as the difference
\bea
d\sigma_e &=& d\sigma_R-d\sigma_L,
\eea
and the parity-violating longitudinal single spin electron asymmetry is defined as
\bea
A_e = \frac{d\sigma_e}{d\sigma_0}.
\eea
We work at tree-level in the partonic cross section and only consider the leading process-independent QED radiation by setting $i=e$,  $j=e$, and $\hat{H}_{ia\to jX}=\delta_{ie}\delta_{aq}\hat{H}_{ia\to jX}$ in Eq.~(\ref{eq:hatH}).
The electron LDF and LFF  are parameterized using the functional form given in Ref.~\cite{Cammarota:2025jyr} at the input scale, $\mu_0=m_c=1.3$ GeV. The  LDFs, LFFs, and PDFs are obtained at $\mu > \mu_0$ by solving the corresponding 
 DGLAP  equations with the evolution kernels at leading order in $\alpha_s$ and $\alpha_{\rm em}$. For simplicity, we choose the LDF and LFF to have the same form, so that $f_{e/e}(\xi,\mu) = D_{e/e}(\xi,\mu) $. Fig.~\ref{fig:ldf} shows  LDFs at the scale $\mu^2=1000$ GeV$^2$ for three parameter choices of the functional form at the input scale, referred to as LF1, LF2, and LF3. We set LF1 as the default distribution~\cite{QiuWatanabe:2026}, 
 and study the impact of different parameterizations through LF2 and LF3.

\begin{table}[]
    \centering
    \begin{tabular}{|c|c|c|}
        \hline
        \hline
        Wilson & Operator  & SMEFT\\
        Coefficient & Label & Operator\\
        \hline
        \hline
       $C_{\ell q}^{(1)}$  & $\mathcal O^{(1)}_{\ell q}$  & $(\bar{L}_L \gamma^\mu L_L)(\bar Q_L \gamma_\mu Q_L)$\\
         \hline
        $C^{(3)}_{\ell q} $ &   $ \mathcal O^{(3)}_{\ell q} $  & $ (\bar L_L \gamma^\mu \tau^I L_L)(\bar Q_L \gamma_\mu \tau^I Q_L) $ \\
        \hline
      $C_{eu}$ &  $ \mathcal O_{eu} $ & $ (\bar e_R \gamma^\mu e_R)(\bar u_R \gamma_\mu u_R) $ \\
      \hline
      $C_{ed}$  & $ \mathcal O_{ed} $ & $ (\bar e_R \gamma^\mu e_R)(\bar d_R \gamma_\mu d_R) $  \\
      \hline
     $C_{\ell u }$ & $ \mathcal O_{\ell u} $ & $ (\bar L_L \gamma^\mu L_L)(\bar u_R \gamma_\mu u_R) $ \\
     \hline
          $C_{\ell d }$ & $ \mathcal O_{\ell d} $ & $ (\bar L_L \gamma^\mu L_L)(\bar d_R \gamma_\mu d_R) $ \\
     \hline
          $C_{qe }$ & $ \mathcal O_{qe} $ & $ (\bar Q_L \gamma^\mu Q_L)(\bar e_R\gamma_\mu e_R) $ \\
     \hline
    \end{tabular}
    \caption{Dimension-6 four-fermion SMEFT operators contributing to the  DIS and Drell-Yan processes.}
    \label{tab:SMEFT}
\end{table}

The calculation of $d\sigma_e$, needed for the longitudinal electron spin asymmetry $A_e$, requires replacing the unpolarized LDF, $f_{e/e}(\xi,\mu)$,  with the polarized LDF, $\Delta f_{e/e}(\xi,\mu)=f_{e/e}^+(\xi,\mu)-f_{e/e}^-(\xi,\mu)$, where $f_{e/e}^+$ and $f_{e/e}^-$ denote the  LDF for an electron with its spin aligned and anti-aligned with the original beam electron, respectively. At leading order in QED perturbation theory, we expect $f_{e/e}^-(\xi,\mu)\approx 0$. 
Thus, for simplicity, we set  $\Delta f_{e/e}(\xi,\mu)\simeq f_{e/e}^+(\xi,\mu)\simeq f_{e/e}(\xi,\mu)$. We note however that the unpolarized and polarized LDFs are inherently non-perturbative distributions due to the possibility of QED radiation producing quark-antiquark pairs that can radiate gluons
nonperturbatively.
Thus, in general one must allow for the possibility that $\Delta f_{e/e}(\xi,\mu)\neq f_{e/e}(\xi,\mu)$. However, we expect this to be a small effect and treat it as negligible in this first analysis.

We parameterize the effects of QED radiation on the cross section and asymmetry as
\bea
\label{eq:devsigma}
d\sigma_0 &=& d\sigma_{0}^{\rm NR}\> (1+\delta^\sigma), \nn \\
A_e &=&A_{e}^{\rm NR}\> (1+\delta^A),
\eea
where $d\sigma_{0}^{\rm NR}$ and $A_{e}^{\rm NR}$ denote the cross section and asymmetry without QED radiation effects included, respectively. The functions, $\delta^\sigma(x_B,Q^2)$ and $\delta^A(x_B,Q^2)$, characterize the relative impact of QED radiation. In Figs.~\ref{fig:deltasigASoLID} and \ref{fig:deltasigAEIC}, we show the functions  $\delta^\sigma$ (top panel) and $\delta^A$ (bottom panel) for a wide range of $(x_B, Q^2)$ values for the SoLID and EIC facilities, respectively. For SoLID, we use a center of mass energy, $\sqrt{s}=4.7$ GeV, corresponding to a 12 GeV electron beam impinging on a proton target. For the EIC, we use $\sqrt{s}=105$ GeV, corresponding to a 10 GeV electron beam colliding with 275 GeV proton beam. The QED radiation effects correspond to the default LF1 parameterization in Fig.~\ref{fig:ldf} with $f_{e/e}(\xi,\mu) = D_{e/e}(\xi,\mu) $.    We see that the impact of QED radiation can be quite large, producing order one shifts in the unpolarized cross section, especially for moderate to large values of $x_B$.  The impact is also typically greater at lower $Q^2$, corresponding to the larger available phase space for QED radiation. By contrast, the impact of QED radiation on the longitudinal electron spin asymmetry is relatively small, producing few percent level shifts. This suggests that analyses based on asymmetries are more robust against the effects of QED radiation compared to those based on cross section measurements. This difference in sensitivity to QED radiation effects between cross sections and asymmetries can be exploited in global fits of LDFs, LFFs, PDFs and the SMEFT operators. 

\begin{figure}[htbp]
    \centering
    \begin{subfigure}{\linewidth}
        \centering
        \includegraphics[width=\linewidth]{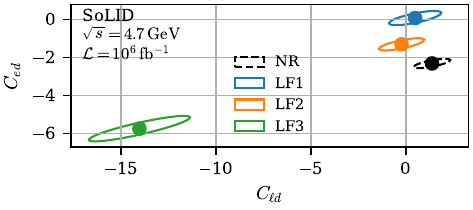}
    \end{subfigure}
    \hfill
    \begin{subfigure}{\linewidth}
        \centering
        \includegraphics[width=\linewidth]{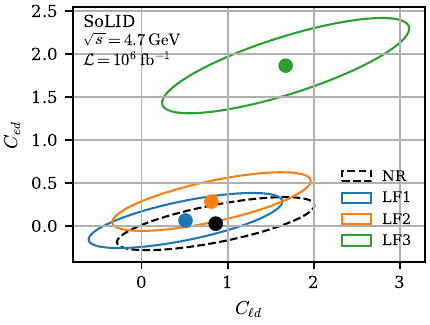}
    \end{subfigure}
    \caption{95\% confidence level contours for the best-fit values of the $( C_{\ell d},C_{ed})$ pair of SMEFT coefficients using 10$^6$ fb$^{-1}$ of pseudo-data for the unpolarized cross section (top) and the longitudinal electron spin asymmetry (bottom) at SoLID. }
    \label{fig:fitSoLID}
\end{figure}

\textit{SMEFT Analysis}:  For experiments with characteristic energy $E\ll \Lambda$, new physics effects at the scale $\Lambda\gg M_W$ can be parameterized by higher dimensional non-renormalizable operators
\begin{eqnarray}
\label{eq:SMEFTLag}
{\cal L}_\mathrm{SMEFT}  = \frac{1}{\Lambda^2}\sum_{r} C_r {\cal O}_r + \cdots  ,
\end{eqnarray}
where the ${\cal O}_r $ denote the dimension-six operators, the $C_r$ denote the Wilson coefficients, and the ellipses denote dimension-eight and higher SMEFT operators. The index $r$ runs over the SMEFT operators listed in Table~\ref{tab:SMEFT} in the basis of SM fields before electroweak symmetry breaking. We characterize the impact of SMEFT operators on the cross section and electron asymmetry as 
\bea
\label{eq:AeCi}
d\sigma_0 &=& d\sigma_{0}^{\rm SM}\> \left(1+\sum_r C_r \>\delta_r^{\sigma, \rm SMEFT}\right), \nn \\
A_e &=& A_e^{\rm SM}\>\left(1+ \sum_r C_r \>\delta_r^{A, \rm SMEFT} \right),
\eea
where $d\sigma_{0}^{\rm SM}$ and $A_e^{\rm SM}$ denote the SM cross section and asymmetry, without contributions from SMEFT operators.  In Eq.~(\ref{eq:AeCi}),  we have expanded the cross section and asymmetry to linear order in the SMEFT Wilson coefficients, $C_r$, as required when working at order $\sim 1/\Lambda^2$ and ignoring contributions from dimension eight SMEFT operators.  The functions $\delta_r^{\sigma, \rm SMEFT}(x_B,Q^2)$ and $\delta_r^{A, \rm SMEFT}(x_B,Q^2)$ are given by $\delta_r^{\sigma, \rm SMEFT}=d\sigma_0^{\rm C_r}/d\sigma_0^{\rm SM}$ and  $\delta_r^{A, \rm SMEFT}=d\sigma_e^{\rm C_r}/d\sigma_e^{\rm SM}-d\sigma_0^{\rm C_r}/d\sigma_0^{\rm SM}$. Here 
 $d\sigma_{0,e}^{\rm SM}$ denote  the unpolarized and polarized cross sections in the SM, respectively.  $d\sigma_{0,e}^{\rm C_r}$ denote the contributions to the unpolarized and polarized cross sections from $r$-th SMEFT operator, respectively, with $C_r$ factored out.  Eq.~(\ref{eq:AeCi}) is applicable both when QED radiation effects are either included or not included. In each of these cases, it is understood that $d\sigma_0^{\rm SM}$, $A_e^{\rm SM}$,  $\delta_r^{\sigma, \rm SMEFT}$, and $\delta_r^{A, \rm SMEFT}$ are correspondingly computed with or without the QED radiation effects.

\begin{figure}[htbp]
    \centering
    \begin{subfigure}{\linewidth}
        \centering
        \includegraphics[width=\linewidth]{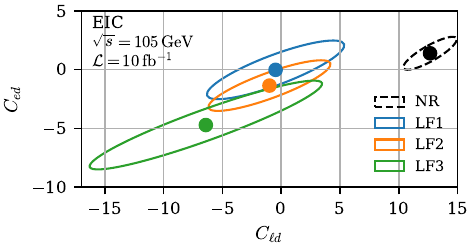}
    \end{subfigure}
    \hfill
    \begin{subfigure}{\linewidth}
        \centering
        \includegraphics[width=\linewidth]{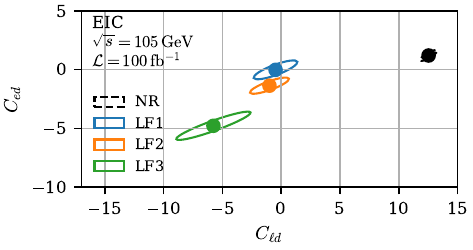}
    \end{subfigure}
    \caption{95\% confidence level contours for the best-fit values of the $(C_{\ell d}, C_{e d})$ pair of SMEFT coefficients using 10 fb$^{-1}$ (top) and 100 fb$^{-1}$ (bottom) of pseudo-data for the unpolarized cross section at EIC.}
     \label{fig:xsecfit}
\end{figure}

\begin{figure}[htbp]
    \centering
    \begin{subfigure}{\linewidth}
        \centering
        \includegraphics[width=\linewidth]{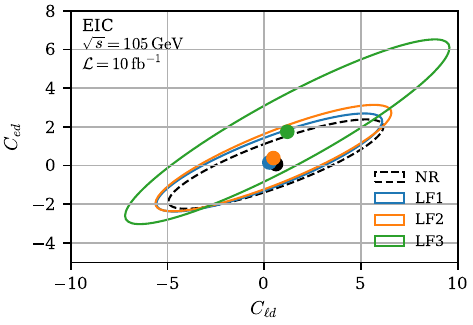}
    \end{subfigure}
    \hfill
    \begin{subfigure}{\linewidth}
        \centering
        \includegraphics[width=\linewidth]{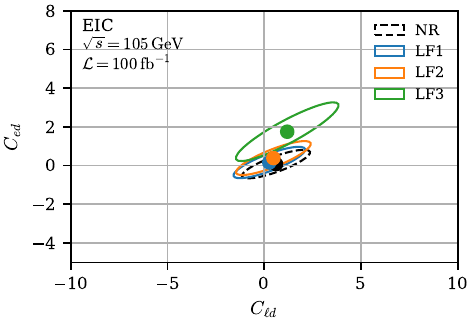}
    \end{subfigure}
    \caption{95\% confidence level contours for the best-fit values of the $(C_{\ell d}, C_{e d})$ pair of SMEFT coefficients using 10 fb$^{-1}$ (top) and 100 fb$^{-1}$ (bottom) of pseudo-data for the longitudinal electron spin asymmetry for the EIC. }
     \label{fig:Aefit}
\end{figure}

\textit{SoLID and EIC Projections}: In this section, we make projections for the impact of QED radiation on the SMEFT constraints that can be extracted from  SoLID and the EIC. We follow the analysis and procedure of Ref.~\cite{Boughezal:2022pmb} to generate pseudo-data  for the EIC and perform a chi-square analysis by comparing theory predictions with the pseudo-data. 
We generate pseudo-data, using the  default LF1 setting to account for QED radiation, for the unpolarized cross section and the longitudinal electron spin asymmetry in each of the $(x_B,Q^2)$ bins shown in Figs.~\ref{fig:deltasigASoLID} and \ref{fig:deltasigAEIC} for SoLID and the EIC, respectively.   The statistical uncertainty is based on an integrated luminosity of 10$^6$ fb$^{-1}$ for SoLID, and two settings of 10fb$^{-1}$ and 100  fb$^{-1}$ for the EIC. The statistical uncertainty is added in quadrature  with a 1\% systematic uncertainty, as well as a fully correlated 1\% electron beam polarization uncertainty. We use an electron beam polarization of $P_e=0.85$ and $P_e=0.80$ for SoLID and the EIC, respectively. 

We perform a chi-square ($\chi^2$) best-fit analysis of the SMEFT Wilson coefficients by comparing the pseudo-data with theory predictions computed with QED radiation turned off (dashed black) and with QED radiation turned on using the LF1 (solid blue), LF2 (solid orange), and LF3 (solid green) parameterizations of the LDF and LFF. For simplicity, we only turn on two of the SMEFT Wilson coefficients in Table~\ref{tab:SMEFT} at a time and set the rest to zero. We have analyzed all  21 distinct pairs (assuming flavor universality) of the 7 SMEFT coefficients. We show results only for the $(C_{\ell d}, C_{e d}) $ pair of SMEFT coefficients, but the results and conclusions are representative of the other pairs of coefficients as well. In our analysis, we have set $\Lambda=1$ TeV. Fig.~\ref{fig:fitSoLID} shows the best-fit values and the corresponding 95\% confidence level contours ($\chi^2<5.991$) using the unpolarized cross section (top) and the electron spin asymmetry (bottom) at SoLID. Figs.~\ref{fig:xsecfit} and  \ref{fig:Aefit} show the corresponding results for the unpolarized cross section and electron spin asymmetry at the EIC, respectively. In each figure, the top and bottom panels correspond to 10 fb$^{-1}$ and 100  fb$^{-1}$ of pseudo-data, respectively.  

Figs.~\ref{fig:fitSoLID},  \ref{fig:xsecfit}, and \ref{fig:Aefit} provide benchmark results that characterize the typical  impact of QED radiation on the extraction of SMEFT Wilson coefficients using unpolarized DIS cross sections and longitudinal electron spin asymmetries. We see that not including QED radiation effects in theory predictions of cross sections can cause large and significant shifts in the extracted best-fit values of the SMEFT coefficients, and affect the size of the 95\% confidence level contours. By contrast,  the corresponding results are much more stable against QED radiation effects when using the longitudinal electron spin asymmetry. We see that the SoLID experiment has sufficient precision and luminosity to clearly distinguish between different LDF and LFF parameterizations using cross section measurements. At the EIC, it requires an integrated luminosity of about 100 fb$^{-1}$. On the other hand,  distinguish between the LF1 and LF2 parameterizations at  SoLID and the EIC is challenging using asymmetries. However, asymmetries can still distinguish between the LF3 and the LF1 or LF2 parameterizations at SoLID, and at the EIC with 100 fb$^{-1}$ of pseudo-data. 

\textit{Conclusions}: Collision-induced QED radiation can produce order one effects in DIS cross sections and few percent level effects in parity-violating longitudinal electron spin asymmetries. Correspondingly, QED radiation effects can significantly bias SMEFT fits derived from DIS cross sections.  By contrast, SMEFT constraints derived from electron asymmetries are relatively more stable against the effects of QED radiation. The joint QCD+QED factorization approach is a rigorous and systematically improvable framework for treating the effects of QED radiation, and serve as an essential ingredient in future SMEFT analyses at SoLID and the EIC.

\textit{Acknowledgements}: S.M. thanks Jefferson Lab for their hospitality and support through the Visiting Faculty Program (VFP) during which part of this work was carried out. The work of J.Q.~and J.Z.~is supported in part by the U.S. Department of Energy, Office of Science, Office of Nuclear Physics under Contract No.~89243126CSC000213.

\bibliographystyle{h-physrev3.bst}
\bibliography{SMEFT}
\end{document}